\documentclass[10pt]{article}

\usepackage{graphicx}
\usepackage{graphics}
\usepackage{amscd}
\usepackage{amsmath}
\usepackage{amsfonts}
\usepackage{amssymb}
\usepackage[english]{babel}

\DeclareMathOperator{\GR}{GR}

\DeclareMathOperator{\w}{w}
\DeclareMathOperator{\hm}{h}

\newcommand{\R}{{\cal R}}
\newcommand{\F}{\mathbb{F}}
\newcommand{\T}{{\cal{T}}(p,m)}

\begin{document}
\title{Some Repeated-Root Constacyclic Codes over Galois Rings$\footnote{
 E-Mail addresses: hwliu@mail.ccnu.edu.cn (H. Liu), m.youcef@mails.ccnu.edu.cn (M. Youcef).}$ }

\author{Hongwei Liu$^1$,~Youcef Maouche$^{1,2}$}

\date{\small
${}^1$School of Mathematics and Statistics, Central China Normal University,Wuhan, Hubei, 430079, China\\
${}^2$Department of Mathematics, University of Sciences and Technology HOUARI BOUMEDIENE, Alger, Algeria\\
}
\maketitle

\leftskip 0.8in
\rightskip 0.8in
\noindent {\bf Abstract.} Codes over Galois rings have been studied extensively during the last three decades.
Negacyclic codes over $\GR(2^a,m)$ of length $2^s$ have been characterized: the ring $\R_2(a,m,-1)= \frac{\GR(2^a,m)[x]}{\langle x^{2^s}+1\rangle}$ is a chain ring. Furthermore, these results have been generalized to $\lambda$-constacyclic codes for any unit $\lambda$ of the form $4z-1$, $z\in \GR(2^a, m)$. In this paper,  we study more general cases and investigate all cases where
$\R_p(a,m,\gamma)= \frac{\GR(p^a,m)[x]}{\langle x^{p^s}-\gamma \rangle}$ is a chain ring. In particular, necessary and sufficient conditions for the ring $\R_p(a,m,\gamma)$ to be a chain ring are obtained. In addition, by using this structure we investigate all $\gamma$-constacyclic codes over $\GR(p^a,m)$ when $\R_p(a,m,\gamma)$ is a chain ring. Necessary and sufficient conditions for the existence of self-orthogonal and self-dual $\gamma$-constacyclic codes are also provided. Among others, for any prime $p$, the structure of $\R_p(a,m,\gamma)=\frac{\GR(p^a,m)[x]}{\langle x^{p^s}-\gamma\rangle}$ is used to establish the Hamming and homogeneous distances of $\gamma$-constacyclic codes.



\leftskip 0.0in
\rightskip 0.0in

\vskip 30pt

\section{Introduction}
\vskip 6pt
The importance of codes over finite rings has been recognized since the 1990s; the works of Nechaev \cite{Nechaev91} and  Hammons et al. \cite{Hammons94}, \cite{calderbank93}  showed that some well-known nonlinear binary codes such as Kerdock and Preparata codes can be constructed from linear codes over $\mathbb{Z}_4$, the ring of integers modulo $4$.
Since then codes over $\mathbb{Z}_4$ in particular, and codes over finite rings in general, have received a great deal of attention.

Constacyclic codes  are a generalization of cyclic codes, and they play a very significant role in the theory of error-correcting codes.
Most works on constacyclic codes over finite rings
concentrate on the situation when the code length  is relatively prime to the characteristic of the underlying ring.
The case where the code length is not relatively prime to the characteristic of the underlying ring yields the so-called repeated-root codes.

\vskip 6pt
Repeated-root constacyclic codes were studied by Castagnoli et al. (cf. \cite{Castagnoli91}),
van Lint (cf.  \cite{Lint91}), and others (see, for example, \cite{tang1997}, \cite{zimmermann}, \cite{nedeloaia2003}), where they showed that repeated-root cyclic codes have a concatenated construction and are asymptotically bad.
Nevertheless, repeated-root codes are optimal in a few cases, which motivates researchers to explore this class of codes further.
Cyclic codes of length $2^s$ over $\mathbb{Z}_4$ were considered by Abualrub and Oehmke in \cite{abualrub},
where such codes were characterized in terms of their sets of generators.
The structure of negacyclic codes of length $2^s$ over $\mathbb{Z}_{2^m}$ was obtained in \cite{DL04}. Moreover, \cite{Blackford03} presented a transform approach to  classify  negacyclic codes of even length over $\mathbb{Z}_4$.

\vskip 6pt
A Galois ring is a Galois extension of the ring $\mathbb{Z}_{p^a}$,
the ring of integers  modulo a prime power $p^a$.
In particular, $\mathbb{Z}_{p^a}$ such as $\mathbb{Z}_4$ is a Galois ring. The class of Galois rings has been used widely as an alphabet for cyclic and negacyclic codes, for instance \cite{Calderbank95}, \cite{minjia1}-\cite{Wan99}, \cite{DL04}, \cite{Babu01}-\cite{Blackford2003}. In 2005, Dinh \cite{dinh2005}  investigated negacyclic codes of length $2^s$ over the Galois ring $\GR(2^a,m)$, and showed that the ring $\R_2(a,m,-1)=\frac{\GR(2^a,m)[x]}{\langle x^{2^s} +1 \rangle}$ is in fact a chain ring.
In 2007, Dinh \cite{dinh2007} computed the Hamming, Lee, homogeneous, and Euclidean distances of all those negacyclic codes over $\mathbb{Z}_{2^a}$. In \cite{liu2016}, the results of Dinh were extended to $\lambda$-constacyclic codes over $\GR(2^a,m)$, for any unit $\lambda$ of the form $4z-1$;
 it was shown that the $\lambda$-constacyclic codes of length $2^s$ over $\GR(2^a,m)$ are precisely the ideals generated by $(x+1)^i$ of the chain ring $\R_2(a,m,\lambda)=\frac{\GR(2^a,m)[x]}{\langle x^{2^s}-\lambda\rangle}$, for $i= 0, 1, \cdots, a2^{s}$. Using this structure, the Hamming, Lee, homogeneous, and Rosenbloom-Tsfasman (RT) distances of all those $\lambda$-constacyclic codes  are obtained
 in the same paper \cite{liu2016}.

\vskip 6pt
The aim of this paper is to study the structures and distances of $\gamma$-constacyclic codes of length $p^s$ over the Galois ring $\GR(p^a,m)$ for any unit $\gamma$ of the form $\zeta_0+p\zeta_1+p^2z$, where $z$ is an arbitrary element of $\GR(p^a,m)$ and $\zeta_0,\zeta_1$ are  nonzero elements of the set $\T$. Here  $\T$ denotes  a complete set of representatives of the cosets $\frac{\GR(p^a,m)}{p\GR(p^a,m)}=\F_{p^m}$ in $\GR(p^a,m)$.
Each unit of this form is called a unit of Type $(1)$. In Section 3, we prove that the cases $-1$ and $4z-1$ are just two special cases of Type $(1)$ when $p=2$. Moreover, for any prime $p$, we show that the ring $\R_p(a,m,\gamma)=\frac{\GR(p^a,m)[x]}{\langle x^{p^s}-\gamma\rangle}$ is a chain ring if and only if $\gamma$ is of Type $(1)$. We also derive the duals of all such $\gamma$-constacyclic codes as well as necessary and sufficient conditions for the existence of self-orthogonal and self-dual $\gamma$-constacyclic codes. In Sections 4 and   5, by using the structure obtained in Section 3,
the Hamming and homogeneous distances of all $\gamma$-constacyclic codes are established respectively.
We conclude this paper with open problems in Section 6.

%

\vskip 30pt
\section{ Preliminaries}

\vskip 6pt
In this paper, all rings under consideration are associative and commutative rings with identity. An ideal $I$ of a ring $R$ is called {\it principal} if it is generated by one element. A ring $R$ is a {\it principal ideal ring} if each of its ideals is principal. We say $R$ is a {\it local} ring if it has a unique maximal ideal. Furthermore, a ring $R$ is called a {\it chain ring} if the set of all ideals of $R$ is linearly ordered under set-theoretic inclusion.

\vskip 6pt
The following proposition is known for the class of finite commutative chain rings (see  \cite[Prop. 2.1]{DL04}).

\vskip 6pt
\noindent {\bf Proposition 2.1.} {\it Let $R$ be a finite commutative ring. Then the following conditions are equivalent:
\begin{description}
\item{$(i)$} $R$ is a local ring and the maximal ideal $M$ of $R$ is principal.
\item{$(ii)$} $R$ is a local principal ideal ring.
\item{$(iii)$} $R$ is a chain ring.
\end{description}}

\vskip 6pt
We have the following well-known properties of chain rings.
\vskip 6pt
\noindent {\bf Proposition 2.2.} {\it Let $R$ be a finite commutative chain ring  with maximal ideal $M=\boldsymbol\langle r \boldsymbol\rangle$. Denote the quotient ring $\bar R =\frac{R}{M}$, and let $\beta$ be the nilpotency of $r$. Then
\begin{description}
\item{$(a)$} There is some prime $p$ and positive integers $k, l$ with $k \geq l$ such that $|R| = p^k, |\bar R|=p^l$, the characteristic of $R$ is powers of $p$ and $\bar R$ is a field.
\item{$(b)$} The ideals of $R$ are $\langle r^i \rangle$, where $i=0,1,\cdots,\beta$, and they are strictly inclusive:
$$R = \langle r^0 \rangle \supsetneq \langle r^1 \rangle \supsetneq \cdots \supsetneq \langle r^{\beta-1} \rangle \supsetneq \langle r^{\beta} \rangle = \langle 0 \rangle.$$
\item{$(c)$} For $i=0,\cdots,\beta$, $|\langle r^i \rangle|=|\bar R|^{\beta-i}$. In particular, $|R|=|\bar R|^{\beta}$, i.e., $k= l\beta$.
\end{description}}

\vskip 6pt
A polynomial in $\mathbb Z_{p^a}[x]$ is called a {\it basic irreducible polynomial} if its reduction modulo $p$ is irreducible in $\mathbb Z_p[x]$. The {\it Galois ring of characteristic $p^a$ and dimension $m$}, denoted by $\GR(p^a,m)$, is the Galois extension of degree $m$ of the ring $\mathbb Z_{p^a}$. Equivalently,
$$\GR(p^a,m) = \frac{\mathbb Z_{p^a}[u]}{\langle h(u) \rangle},$$
where $h(u)$ is a monic basic irreducible polynomial of degree $m$ in $\mathbb Z_{p^a}[u]$.
Note that if $a=1$, then $\GR(p,m)=\F_{p^m}$, and if $m=1$  then $\GR(p^a,1)=\mathbb Z_{p^a}$. We list some well-known facts about Galois rings (cf. \cite{M74,HP03,PH98}), which will be used throughout this paper.

\vskip 6pt
\noindent {\bf Proposition 2.3.} {\it Let $\GR(p^a,m)= \frac{\mathbb Z_{p^a}[u]}{\langle h(u) \rangle}$ be a Galois ring. Then the following hold:
\begin{description}
\item{$(i)$} Each ideal of $\GR(p^a,m)$ is of the form $\langle p^k \rangle=p^k\GR(p^a,m)$, for $0 \leqslant k \leqslant a$. In particular, $\GR(p^a,m)$ is a chain ring with maximal ideal $\langle p \rangle = p\GR(p^a,m)$  and residue field $\F_{p^m}$.

\item{$(ii)$} For $0 \leqslant i \leqslant a$, $|p^i\GR(p^a,m)|=p^{m(a-i)}$.

\item{$(iii)$} Each element of $\GR(p^a,m)$ can be represented as $vp^k$, where $v$ is a unit and $0 \leqslant k \leqslant a$. In this representation $k$ is unique and $v$ is unique modulo $\langle p^{a-k} \rangle$.

\item{$(iv)$} $h(u)$ has a root $\zeta$ in $\GR(p^a,m)$, which is also a primitive $(p^m-1)$th root of unity. The set
$${\cal T}(p,m)=\big\{0, 1, \zeta, \zeta^2, \cdots, \zeta^{p^m-2}\big\}$$
is a complete set of representatives of the cosets $\frac{\GR(p^a,m)}{p\GR(p^a,m)}=\F_{p^m}$ in $\GR(p^a,m)$. Each element $r \in \GR(p^a,m)$ can be written uniquely as
$$r = \zeta_0 + p\zeta_1 + \cdots + p^{a-1}\zeta_{a-1}$$
with $\zeta_i \in {\cal T}(p,m)$, $0 \leqslant i \leqslant a-1$.

\item{$(v)$} For $0\leq i < j\leq p^m-2$, all $\zeta^i-\zeta^j$ are units of $\GR(p^a,m)$.
\end{description}}

\vskip 6pt
For a finite ring $R$, consider the set $R^n$ of $n$-tuples of elements from $R$ as a module over $R$. Any nonempty subset $C \subseteq R^n$ is called a {\it code of length $n$} over $R$, and the code $C$ is {\it linear} if in addition, $C$ is an $R$-submodule of $R^n$. Let $\lambda$ be a unit of the ring $R$, then the {\it $\lambda$-constacyclic} {\it shift} $\tau_{\lambda}$ on $R^n$ is the shift
$$\tau_{\lambda}\big(x_0,x_1,\cdots,x_{n-1}\big)=\big(\lambda x_{n-1},x_0,x_1,\cdots,x_{n-2}\big),$$
and a code $C$ is said to be {\it $\lambda$-constacyclic} if $\tau_{\lambda}(C)=C$, i.e., if $C$ is closed under the $\lambda$-constacyclic shift $\tau_{\lambda}$. In light of this definition, when $\lambda=1$, $\lambda$-constacyclic codes are just cyclic codes, and when $\lambda=-1$, $\lambda$-constacyclic codes are called negacyclic codes.

\vskip 6pt
Each codeword $c=(c_0,c_1,\cdots,c_{n-1})$ of a code $C$ is customarily identified with its polynomial representation $c(x)=c_0+c_1x+\cdots+c_{n-1}x^{n-1}$, and the code $C$ is in turn identified with the set of all polynomial representations of its codewords. Then in the ring $\frac{R[x]}{\langle x^n-\lambda \rangle}$, $xc(x)$ corresponds to a $\lambda$-constacyclic shift of $c(x)$. From that, the following well-known fact is straightforward:

\vskip 6pt
\noindent
{\bf Proposition 2.4.} {\it A linear code $C$ of length $n$ over $R$ is $\lambda$-constacyclic if and only if $C$ is an ideal of $\frac{R[x]}{\langle x^n-\lambda \rangle}$.}

\vskip 6pt
Given $n$-tuples $x = (x_0, x_1, \cdots, x_{n-1}), y = (y_0, y_1, \cdots, y_{n-1}) \in R^n$, their {\it inner product} is defined as usual
$$\langle x, y\rangle = x_0y_0 + x_1y_1 + \cdots +x_{n-1}y_{n-1}\in R.$$
Two $n$-tuples $x, y$ are called {\it orthogonal} if $\langle x, y\rangle= 0$. For a linear code $C$ over $R$, its {\it dual code} $C^{\perp}$ is the set of $n$-tuples over $R$ that are orthogonal to all codewords of $C$, i.e.,
$$C^{\perp} = \big\{x \, \big| \, \langle x, y\rangle = 0, \forall y \in C \big\}.$$
A code $C$ is called {\it self-orthogonal} if $C \subseteq C^{\perp}$, and it is called {\it self-dual} if $C = C^{\perp}$. The following result is well-known (cf. \cite{DL04}).

\vskip 6pt
\noindent {\bf Proposition 2.5.} {\it Let $R$ be a finite chain ring of size $p^{\alpha}$. The number of codewords in any linear code $C$ of length $n$ over $R$ is $p^{k}$, for some integer $k$, $0 \leq k \leq \alpha n$. Moreover, the dual code $C^{\perp}$ has $p^{\alpha n-k}$ codewords, so that $|C| \cdot |C^{\perp}| = |R|^n$.}

\vskip 6pt
Note that the dual of a cyclic code is a cyclic code, and the dual of a negacyclic code is a negacyclic code. In general, we have the following implication of the dual of a $\lambda$-constacyclic code.

\vskip 6pt
\noindent {\bf Proposition 2.6.} {\it The dual of a $\lambda$-constacyclic code is a $\lambda^{-1}$-constacyclic code.}

\section{Constacyclic codes of length $p^s$ over $\GR(p^a,m)$ }

As  mentioned in Section 2, there exists a primitive $(p^m-1)$th root of unity $\zeta$ such that the set
$$\T=\lbrace 0,1,\zeta,\zeta^2,\cdots,\zeta^{p^m-2} \rbrace$$

\noindent is a complete set of representatives of the cosets $\frac{\GR(p^a,m)}{p\GR(p^a,m)}=\F_{p^m}$ in $\GR(p^a,m)$. Each element $r \in \GR(p^a,m)$ can be written uniquely as
$$r =\zeta_0 +p\zeta_1 +p^2\zeta_2 +\cdots+ p^{a-1}\zeta_{a-1}$$
\noindent with $\zeta_i \in \T$, $0 \leq i \leq a-1 $. To simplify notations, we will say that an element $\gamma \in \GR(p^a,m)$ is of {\em Type $(0)$} if it
has the form $\gamma = \zeta_0 +p^2\zeta_2 +\cdots+ p^{a-1}\zeta_{a-1}=\zeta_0 +p^2z$, with $\zeta_0\not =0$, and $\gamma$ is said to be of {\em Type $(1)$} if it is of the form $\gamma = \zeta_0 +p\zeta_1 +p^2\zeta_2 +\cdots+ p^{a-1}\zeta_{a-1}=\zeta_0 +p\zeta_1 +p^2z$, for $ \zeta_0 \not = 0 \not = \zeta_1$ and $z \in \GR(p^a,m)$. Clearly, the elements of Type $(0)$ and Type $(1)$ are invertible in $\GR(p^a,m)$. Furthermore, the sets of Type $(0)$ and Type $(1)$ form a partition of the set of all units of $\GR(p^a,m)$ when $a \geq 2$. We call a $\gamma$-constacyclic code is of Type $(0)$ (resp. Type $(1)$) if the unit $\gamma$ is of Type $(0)$ (resp. Type $(1)$). By Proposition 2.4, $\gamma$-constacyclic codes of length $p^s$ over $\GR(p^a,m)$ are exactly the ideals of the ambient ring
$$ \R_p(a,m,\gamma)=\frac{\GR(p^a,m)[x]}{\left\langle x^{p^s}-\gamma \right\rangle}.$$

\vskip 6pt
\noindent {\bf Proposition 3.1.} { Let $b$ and $\lambda$ be two units of $\GR(p^a, m)$. For any positive integer $n$, there exist polynomials $\alpha_n(x),\beta_n(x),\theta_n(x) \in \mathbb{Z}[x]$, such that
\begin{itemize}
\item If $p=2$, then $(x+b)^{2^n}=x^{2^n} + b^{2^n} + 2\alpha_n(x)=x^{2^n} + b^{2^n} + 2((bx)^{2^{n-1}}+2\beta_n(x))$. Moreover, $\alpha_n(x)$ is invertible in $\R_2(a,m,\lambda)$.

\item If $p$ is odd, then $(x+b)^{p^n} = x^{p^n} + b^{p^n} + p(x+b)\theta_n(x)$.

\end{itemize}

}
\vskip 6pt
{\it Proof.} We prove this proposition by induction on $n$. If $p=2$ and $n=1$, then $(x+b)^2=x^2+b^2+2bx$, $\alpha_1(x)=bx $ and $\beta_1(x)=0$. Obviously, $\alpha_1(x)=bx$ is a unit in $\R_2(a,m,\lambda)$. Assume $n > 1$ and the conclusion is true for all positive integers less than $n$. Then
\begin{align*}
(x+b)^{2^n} &= ((x+b)^{2^{n-1}})^2=(x^{2^{n-1}} + b^{2^{n-1}} + 2\alpha_{n-1}(x))^2\\
    &= x^{2^n} + b^{2^n} + 4\alpha^2_{n-1}(x)+4b^{2^{n-1}}\alpha_{n-1}(x)+4x^{2^{n-1}}\alpha_{n-1}(x)+2(bx)^{2^{n-1}}\\
    &= x^{2^n} + b^{2^n}+2\alpha_n(x),
\end{align*}
where $\alpha_n(x)=(bx)^{2^{n-1}}+2\beta_n(x)$ and $\beta_n(x)=\alpha^2_{n-1}(x)+b^{2^{n-1}}\alpha_{n-1}(x)+x^{2^{n-1}}\alpha_{n-1}(x)$.
We know that both of $x$ and $b$ are invertible in $\R_2(a,m,\lambda)$, and so $(bx)^{2^{n-1}}$ is also invertible in $\R_2(a,m,\lambda)$. As $2$ is nilpotent in $\R_2(a,m,\lambda)$, the proof is completed for $p=2$.

Now suppose $p$ is odd. Let $m$ be a positive integer, then we have
\begin{align*}
(x^{p^{m-1}}+b^{p^{m-1}})^p&=x^{p^m}+b^{p^m}+\sum_{i=1}^{p-1} \binom{p}{i} (b^{p^{m-1}})^i(x^{p^{m-1}})^{p-i}\\
  &=x^{p^m}+b^{p^m}+\sum_{i=1}^{\frac{p-1}{2}}\left\lbrace \binom{p}{i} (x^{p^{m-1}})^{p-i}(b^{{p^{m-1}}})^{i} + \binom{p}{p-i} (x^{{p^{m-1}}})^{i}(b^{{p^{m-1}}})^{p-i} \right\rbrace\\
&=x^{p^m}+b^{p^m}+\sum_{i=1}^{\frac{p-1}{2}} \binom{p}{i}
     b^{ip^{m-1}} x^{ip^{m-1}} \left\lbrace  x^{p^{m-1}(p-2i)}+b^{p^{m-1}(p-2i)} \right\rbrace.
\end{align*}
Clearly, $p^{m-1}(p-2i)$ is odd, thus there exist polynomials $\beta_i^\prime(x)\in \mathbb{Z}[x]$, $0 \leq i \leq \frac{p-1}{2}$, such that $x^{p^{m-1}(p-2i)}+b^{p^{m-1}(p-2i)}=(x+b)\beta_i^\prime(x)$. Then
\begin{align*}
(x^{p^{m-1}}+b^{p^{m-1}})^p&=x^{p^m}+b^{p^m}+\sum_{i=1}^{\frac{p-1}{2}} \binom{p}{i} b^{ip^{m-1}} x^{ip^{m-1}}(x+b)\beta_i^\prime(x) \\
   &=x^{p^m}+b^{p^m}+p(x+b)\sum_{i=1}^{\frac{p-1}{2}} \frac{\binom{p}{i}}{p}b^{ip^{m-1}} x^{ip^{m-1}}\beta_i^\prime(x).
\end{align*}
Hence, we have
\begin{equation}
(x^{p^{m-1}}+b^{p^{m-1}})^p=x^{p^m}+b^{p^m}+p(x+b)\beta_m^\prime(x),
\end{equation}
where
$$ \beta_m^\prime(x)= \sum_{i=1}^{\frac{p-1}{2}} \frac{\binom{p}{i}}{p}b^{ip^{m-1}} x^{ip^{m-1}}\beta_i^\prime(x).$$
Plugging in $m=1$ yields that the conclusion is true for $n=1$. Assume $n > 1$ and the conclusion is true for all positive integers less than $n$. Then
 \begin{align*}
(x+b)^{p^n} &=((x+b)^{p^{n-1}})^p =(x^{p^{n-1}}+b^{p^{n-1}}+p(x+b)\alpha_{n-1}(x))^p \\
		&=(x^{p^{n-1}}+b^{p^{n-1}})^p+\sum_{i=1}^{p}\binom{p}{i}(x^{p^{n-1}}+b^{p^{n-1}})^{p-i}(p(x+b)\alpha_{n-1}(x))^i\\
		&=(x^{p^{n-1}}+b^{p^{n-1}})^p+p(x+b)t(x),
\end{align*}
where
$$ t(x)= \sum_{i=1}^{p} \binom{p}{i}(x^{p^{n-1}}+b^{p^{n-1}})^{p-i} \frac{(p(x+b)\alpha_{n-1}(x))^i}{p(x+b)}.$$
By using Equation (1) and inductive hypothesis, we get
\begin{align*}
(x+b)^{p^n} &=x^{p^n}+b^{p^n}+p(x+b)\beta_n^\prime(x)+p(x+b)t(x)=x^{p^n}+b^{p^n}+p(x+b)\theta_n(x),
\end{align*}
where $\theta_n(x)=\beta_n^\prime(x)+t(x)$. The proof is complete. $\square$

\vskip 4pt
Note that the ring $\R_p(a,m,\lambda)$ is a local ring, and hence in $\R_p(a,m,\lambda)$ the sum of two noninvertible elements is noninvertible, and the sum of a noninvertible element and an invertible element is invertible.

\vskip 6pt
\noindent {\bf Lemma 3.2.}{ Let $\lambda$ be a unit of Type $(1)$ of $\GR(p^a,m)$, i.e., $\lambda = \zeta_0+p\zeta_1+p^2z$ with some $z\in \GR(p^a,m)$ and  $\zeta_0$, $\zeta_1$ nonzero elements of $ \mathcal{T}(p,m)$. Then there exists an invertible element $\alpha$ in $\T$, such that $\langle(x-\alpha)^{p^s}\rangle = \langle p\rangle$ in $\R_p(a,m,\lambda)$, and the element $x-\alpha$ is nilpotent with nilpotency $ap^s$. }

\vskip 6pt
{\it Proof.} We have that $\T\backslash\lbrace0\rbrace \simeq \F^*_{p^m}$, and $\T\backslash\lbrace0\rbrace$ is generated by $\zeta$. Note that $\gcd(p^s,\vert\F^*_{p^m}\vert)=\gcd(p^s,p^m-1)=1$. This implies that $\zeta^{p^s}$ is also a generator of $\mathcal{T}(p,m)\backslash\lbrace0\rbrace $. Therefore, there exists integer $i$, $0 \leq i \leq p^m-1$ such that $\zeta^{ip^s}=\zeta_0$. Let $\alpha=\zeta^{i}$ then $\alpha^{p^s}=\zeta_0$. If $p=2$, by Proposition 3.1 we have
\begin{align*}
(x-\alpha)^{2^s} &= x^{2^s} + (- \alpha)^{2^s}+2\alpha_s(x) \\
                  &= \lambda + \alpha^{2^s}+2[(-x\alpha)^{2^{s-1}} +2\beta_s(x)]\\
                  &=\zeta_0+2\zeta_1+4z+\zeta_0+2[(x\alpha)^{2^{s-1}} +2\beta_s(x)]\\
                  &=2[(x\alpha)^{2^{s-1}}+\zeta_0+\zeta_1+2(\beta_s(x)+z)].
\end{align*}
To complete our proof we first need to prove that $(x\alpha)^{2^{s-1}}+\zeta_0$ is noninvertible. Now suppose to the contrary that $(x\alpha)^{2^{s-1}}+\zeta_0$ is invertible in ${\cal R}_2(a,m,\lambda)$, then $$(x\alpha)^{2^{s-1}}-\zeta_0=[(x\alpha)^{2^{s-1}}+\zeta_0]-2\zeta_0 ,$$
is invertible in ${\cal R}_2(a,m,\lambda)$, which implies that $((x\alpha)^{2^{s-1}}-\zeta_0)((x\alpha)^{2^{s-1}}+\zeta_0)=(x\alpha)^{2^{s}}-\zeta_0^2$ is also invertible in ${\cal R}_2(a,m,\lambda)$. This is a contradiction, since in  ${\cal R}_2(a,m,\lambda)$

$$(x\alpha)^{2^{s}}-\zeta_0^2=\lambda \alpha^{2^s}-\zeta_0^2=(\zeta_0+2\zeta_1+4z)\zeta_0-\zeta_0^2=2(\zeta_0\zeta_1+2\zeta_0 z).$$
Therefore, $(x\alpha)^{2^{s-1}}+\zeta_0$ is noninvertible in ${\cal R}_2(a,m,\lambda)$.
Clearly, $2(\beta_n(x)+z)$ is noninvertible in ${\cal R}_2(a,m,\lambda)$, which implies that $\zeta_1+((x\alpha)^{2^{s-1}}+\zeta_0)+2(\beta_n(x)+z$) is invertible. Hence, $\langle(x-\alpha)^{2^s}\rangle = \langle2\rangle$, and $x-\alpha$ has nilpotency $a2^s$.

If $p$ is odd, by using Proposition 3.1 again,
\begin{align*}
(x-\alpha)^{p^s} &= x^{p^s} + (-\alpha)^{p^s} + p(x-\alpha)\alpha_s(x) \\
				 &= \lambda - \alpha^{p^s} + p(x-\alpha)\alpha_s(x)\\	
				 &= \zeta_0+p\zeta_1+p^2z-\zeta_0+p(x-\alpha)\alpha_s(x)\\
				 &=p(\zeta_1+pz+(x-\alpha)\alpha_s(x)).
\end{align*}
Because $p$ is nilpotent in $\GR(p^a,m)$, $x-\alpha$ is also nilpotent. It follows that $pz+(x-\alpha)\alpha_s(x)$ is a noninvertible element in $\R_p(a,m,\lambda)$, which implies that $\zeta_1+pz+(x-\alpha)\alpha_s(x)$ is invertible. Hence, $\langle(x-\alpha)^{p^s}\rangle = \langle p\rangle$, and $x-\alpha$ has nilpotency $ap^s$. $\square $

\vskip 6pt
\noindent {\bf Theorem 3.3.}{ Let $\lambda=\zeta_0+p\zeta_1+p^2z \in \GR(p^a,m)$ be a unit of Type $(1)$. Then the ring $\R_p(a, m,\lambda)$ is a chain ring with maximal ideal $\langle x - \alpha\rangle$, where $\alpha^{p^s}=\zeta_0$. The $\lambda$-constacyclic codes of length $p^s$ over  $\GR(p^a,m)$ are precisely the ideals $\langle(x -\alpha)^i\rangle$ of the ring $\R_p(a,m,\lambda)$, where $0 \leq i \leq ap^s$. Each $\lambda$-constacyclic code $\langle(x - \alpha)^i\rangle$ has exactly $p^{m(p^sa-i)}$ codewords.}

\vskip 6pt
{\it Proof.} Let $f(x)\in \R_p(a,m,\lambda)$, then $f(x)$ can be expressed as
$$f(x)=b_0+b_1(x-\alpha)+b_2(x-\alpha)^2+\cdots+b_{p^{s}-1}(x-\alpha)^{p^{s}-1},$$
where $b_i \in \GR(p^a,m)$. Clearly,
$$b_1(x-\alpha)+b_2(x-\alpha)^2+\cdots+b_{p^{s}-1}(x-\alpha)^{p^{s}-1}$$
 is noninvertible in $\R_p(a,m,\lambda)$. Since $\R_p(a,m,\lambda)$ is a local ring, $f(x)$ is noninvertible if and only if $b_0\in p\GR(p^a,m)$. Moreover, by Lemma 3.2, $p \in \langle(x - \alpha)^{p^s}\rangle \subsetneq \langle x - \alpha\rangle$. Hence, $\langle x-\alpha\rangle$ is the set of all noninvertible elements of $\R_p(a,m,\lambda)$, which implies that $\R_p(a,m,\lambda)$ is a chain ring with maximal ideal $\langle x-\alpha\rangle$. By Lemma 3.2 again, the nilpotency of $ x-\alpha$ is $ap^{s}$, so the ideals of $\R_p(a,m,\lambda)$ are $\langle(x - \alpha)^i\rangle$, $0 \leq i \leq ap^{s}$. The rest of the theorem follows readily from the fact that $\lambda$-constacyclic codes of length $p^s$ over $\GR(p^a,m)$ are ideals of the chain ring $\R_p(a,m,\lambda)$. $\square$

\vskip 6pt
\noindent {\bf Lemma 3.4.}{ Let $\gamma_1 = \zeta_{00}+p\zeta_{01}+p^2z_1$ and $\gamma_2 = \zeta_{10}+p\zeta_{11}+p^2z_2$ be two units of Type $(1)$. Let $\gamma_3 =1+p^2z_3$ and $\gamma_4 = 1+p^2z_4$ be two units of Type $(0)$. Let $a_0 \ge 2$ be the smallest integer such that $2^{a_0} \ge a$, i.e., $p^{2^{a_0}}=0$ in $\GR(p^a,m)$. Then
\begin{itemize}

\item $\gamma_1\gamma_3$ is of Type $(1)$, i.e., the product of a unit of Type $(1)$ and a unit of Type $(0)$ is a unit of Type $(1)$.
\item $\gamma_3\gamma_4$ is of Type $(0)$, i.e., the product of two units of Type $(0)$ is a unit of Type $(0)$.

\item$\gamma_1^{-1}=\zeta_{00}^{-1}(1-p(\zeta_{00}^{-1}\zeta_{01}+p\zeta_{00}^{-1}z_1))
\prod_{j=1}^{a_0-1}[1+p^{2^j}(\zeta_{00}^{-1}\zeta_{01}+p\zeta_{00}^{-1}z_1)^{2^j}]$   is of Type $(1)$, i.e., the inverse of a unit
of Type $(1)$ is a unit of Type $(1)$.

\item $\gamma_3^{-1}=(1-p^2z_3)\prod_{j=1}^{a_0-1}[1+(p^2z_3)^{2^j}]$ is of Type $(0)$, i.e., the inverse of a unit of Type $(0)$ is a unit of Type $(0)$.
\end{itemize}
}
\vskip 6pt
{\it Proof.}
The first and the second statements follow readily. For the third statement, observe that
\begin{eqnarray*}
&&(1-p(\zeta_{00}^{-1}\zeta_{01}+p\zeta_{00}^{-1}z_1))(1+p(\zeta_{00}^{-1}\zeta_{01}+p\zeta_{00}^{-1}z_1))\prod_{j=1}^{a_0-1}[1+p^{2^j}(\zeta_{00}^{-1}\zeta_{01}+p\zeta_{00}^{-1}z_1)^{2^j}]\\
&&=(1-p^2(\zeta_{00}^{-1}\zeta_{01}+p\zeta_{00}^{-1}z_1)^2)\prod_{j=1}^{a_0-1}[1+p^{2^j}(\zeta_{00}^{-1}\zeta_{01}+p\zeta_{00}^{-1}z_1)^{2^j}]=1-p^{2^{a_0}}(\zeta_{00}^{-1}\zeta_{01}+p\zeta_{00}^{-1}z_1)^{2^{a_0}}=1.
\end{eqnarray*}
Therefore,
$$\gamma_1^{-1}\zeta_{00}=(1-p(\zeta_{00}^{-1}\zeta_{01}+p\zeta_{00}^{-1}z_1))\prod_{j=1}^{a_0-1}[1+p^{2^j}(\zeta_{00}^{-1}\zeta_{01}+p\zeta_{00}^{-1}z_1)^{2^j}].$$
To complete the proof,  it suffices to show that $\zeta_{00}^{-1}-p(\zeta_{01}\zeta_{00}^{-2}+p\zeta_{00}^{-2}z_1)$ is of Type $(1)$. Since $-\zeta_{01}\zeta_{00}^{-2}$ is invertible in $\GR(p^a,m)$, $-\zeta_{01}\zeta_{00}^{-2}=\zeta^\prime+pz,$ where $0\ne \zeta'\in\T$. It implies that
$$ \zeta_{00}^{-1}-p(\zeta_{01}\zeta_{00}^{-2}+p\zeta_{00}^{-2}z_1)= \zeta_{00}^{-1}+p\zeta^\prime+p^2z^\prime,$$
where $z^\prime=-\zeta_{00}^{-2}z_1\in \GR(p^a,m)$. Hence $\zeta_{00}^{-1}-p(\zeta_{01}\zeta_{00}^{-2}+p\zeta_{00}^{-2}z_1)$ is of Type $(1) $. Note that for $1\leq j\leq a-1$,  $1+p^{2^j}(\zeta_{01}^{-1}\zeta_{01}+p\zeta_{01}^{-1}z_1)^{2^j}$ is of Type $(0)$. The rest  follows from the first two statements.

The proof of the fourth statement is similar to that of the third statement.   $\square$

\vskip 6pt
\noindent {\bf Proposition 3.5.}{ Let $\gamma=\zeta_0+p\zeta_1+p^2z \in \GR(p^a,m)$ be a unit of Type $(1)$, and let $C=\langle (x-\alpha)^i\rangle \subseteq \R_p(a,m,\gamma)$ be a $\gamma$-constacyclic code of length $p^s$ over $\GR(p^a,m)$, for some $i \in \lbrace 0,1,\cdots, ap^s\rbrace$, where $\alpha^{p^s}=\zeta_0$. The dual of $C$ is a $\gamma^{-1}$-constacyclic code of length $p^s$ over $\GR(p^a,m)$, and $C^\perp =\langle(x -\alpha^{-1})^{ap^s-i}\rangle \subseteq \R_p(a,m,\gamma^{-1})$  which contains precisely $p^{mi}$ codewords.
}
\vskip 6pt
{\it Proof.}
By Proposition 2.6, $C^\perp$ is a $\gamma^{-1}$-constacyclic code of length $p^s$ over $\GR(p^a, m)$. By Lemma 3.4, $\gamma^{-1}=\zeta_0^{-1}+p\zeta^\prime+p^2z^\prime$ is also a unit of Type $(1)$. Thus, Theorem 3.3 is applicable for $C^\perp$ and $\R_p(a, m, \gamma^{-1})$. Observe that $ (\alpha^{-1})^{p^s}=\zeta_0^{-1} $. Hence, $C^\perp$ is an ideal of the form $\langle (x-\alpha^{-1})^j \rangle \subseteq \R_p(a, m,\gamma^{-1})$, where $0 \leq j \leq ap^s$. On the other hand, by Proposition 2.5,
$$\vert C\vert \cdot  \vert C^\perp \vert=\vert \GR(p^a,m)\vert^{p^s}=p^{p^sam},$$
which implies that
$$ \vert C^\perp \vert=\frac{p^{p^sam}}{\vert C\vert}=\frac{p^{p^sam}}{p^{m(p^sa-i)}}=p^{mi}.$$
Therefore, $C^\perp$ must be the ideal $\langle(x-\alpha^{-1})^{p^sa-i}\rangle $ of $\R_p(a, m, \gamma^{-1})$. $\square$

\vskip 6pt
The following definition was introduced in \cite{dinh2017}.

\vskip 6pt
\noindent {\bf Definition 3.6.}{ Let $C$ be a linear code of length $n$ over a finite ring $R$ such that $C$ is both $\alpha$- and $\beta$-constacyclic, for distinct units $\alpha$, $\beta$ of $R$. Then $C$ is called a multi-constacyclic code, or more specifically, an $[\alpha,\beta]$-multi-constacyclic code.}

\vskip 6pt
It is known that  a code $C$ of length $n$ over a finite field $\F$ is a multi-constacyclic code  if and only if $C=\lbrace 0 \rbrace$ or $C=\F^n$. There are non-trivial multi-constacyclic codes over a finite ring $R$.

\vskip 6pt
\noindent {\bf Proposition 3.7.}{ Let $\lambda_1=\zeta_0+p\zeta_1+p^2z_1$, $\lambda_2=\zeta_0+p\zeta^\prime_1+p^2z_2$ be two distinct units of Type $(1)$, and let $C=\langle (x-\alpha)^i \rangle \subseteq \R_p(a,m,\lambda_1)$ be a $\lambda_1$-constacyclic code of length $p^s$ over $\GR(p^a,m)$. Then $C$ is also a $\lambda_2$-constacyclic code, i.e., $C$ is a $[\lambda_1, \lambda_2]$-multi-constacyclic code.

}

\vskip 6pt
{\it Proof.}
By the division algorithm, there exist nonnegative integers $j, t$ such that $i=tp^s+j$, $0 \leq j <p^s$. Using Lemma 3.2, then we have
$$ C=\langle (x-\alpha)^i \rangle=\langle (x-\alpha)^{tp^s}(x-\alpha)^j \rangle=\langle p^t(x-\alpha)^j \rangle.$$
Let $c$ be an arbitrary codeword of $C$, then $c$ has the form $c=p^t(c_0,c_1,\cdots,c_{p^s-1})$. Note that $C$ is a $\lambda_1$-constacyclic code, and we have
\begin{align*}
p^t(\lambda_1 c_{p^s-1},c_0,\cdots,c_{p^s-2})&=p^t((\zeta_0+p\zeta_1+p^2z_1) c_{p^s-1},c_0,\cdots,c_{p^s-2})\\
	&=p^t(\zeta_0 c_{p^s-1},c_0,\cdots,c_{p^s-2})+p^{t+1}((\zeta_1+pz_1)c_{p^s-1},0,\cdots,0) \in C.
\end{align*}
On the other hand,
$$p^{t+1}\in \langle p^{t+1}\rangle=\langle (x-\alpha)^{(t+1)p^s}\rangle  \subseteq\langle (x-\alpha)^{tp^s+j}\rangle=C.$$
This implies that $(p^{t+1},0,\cdots,0)\in C$. Since $C$ is a linear code and $p^{t+1}(\zeta_1+pz_1)c_{p^s-1},p^{t+1}(\zeta_1^\prime+pz_1^\prime)c_{p^s-1}\in \GR(p^a,m) $, we have
 $$p^{t+1}((\zeta_1+pz_1)c_{p^s-1},0,\cdots,0) \mbox{ and }  p^{t+1}((\zeta^\prime_1+pz_2)c_{p^s-1},0,\cdots,0) \in C,$$
which yields that
\begin{align*}
 p^t(\lambda_2c_{p^s-1},c_0,\cdots,c_{p^s-2})=p^t(\zeta_0c_{p^s-1},c_0,\cdots,c_{p^s-2})+p^{t+1}((\zeta^\prime_1+pz_2)c_{p^s-1},0,\cdots,0) \in C.
\end{align*}
Thus, $C$ is also a $\lambda_2$-constacyclic code. $\square$

\vskip 6pt
\noindent {\bf Corollary 3.8.}{ Let $\lambda_1=\zeta_0+p\zeta_1+p^2z_1$ and $\lambda_2=\zeta_0+p\zeta^\prime_1+p^2z_2$ be two units of Type $(1)$.
Let $C=\langle (x-\alpha)^i \rangle \subseteq \R_p(a,m,\lambda_1)$ be a $\lambda_1$-constacyclic code of length $p^s$ over $\GR(p^a,m)$. Then $C$ is also the ideal $\langle (x-\alpha)^i \rangle $ of the ring $\R_p(a,m,\lambda_2)$,
 i.e., let $c(x) \in \GR(p^a,m)[x]$ be a polynomial of degree less than $p^s$, then there exists a polynomial $g(x)\in \GR(p^a,m)[x]$  such that $c(x)\equiv g(x)(x-\alpha)^i\mod (x^{p^s}-\lambda_1) $ if and only if there exists a polynomial $g^\prime(x)\in \GR(p^a,m)[x]$  such that $c(x)\equiv g^\prime(x)(x-\alpha)^i\mod (x^{p^s}-\lambda_2) $.}
\vskip 6pt
{\it Proof.} By Proposition 3.7, $C$ is also a $\lambda_2$-constacyclic code which contains $p^{m(ap^s-i)}$ codewords. By Proposition 2.4, $C$ is an ideal of the ring $\R_p(a,m,\lambda_2)$, because $\lambda_2$ is of Type $(1)$ and $\alpha^{p^s}=\zeta_0$. Thus, Theorem 3.3 is applicable for $C$ and $\R_p(a,m,\lambda_2)$. Hence, $C$ is the ideal $\langle (x-\alpha)^i \rangle $ of the ring $\R_p(a,m,\lambda_2)$. $\square$

\vskip 6pt
\noindent {\bf Remark 3.9.}{ Corollary 3.8 gives us very important information about $\lambda$-constacyclic codes over $\GR(p^a,m)$, where $\lambda$ is a unit of Type $(1)$. This corollary shows that the $\lambda$-constacyclic codes depend on $\zeta_0$ only, which means that there exist just $p^m-1$ different codes of length $p^s$ over $\GR(p^a,m)$ of Type $(1)$. Moreover, in Section 4, we will show that those codes are similar to $\bar{\lambda}$-constacyclic codes of length $p^s$ over $\F_{p^m}$.}

\vskip 6pt
\noindent {\bf Theorem 3.10.} { Let $\gamma=\zeta_0+p\zeta_1+p^2z \in \GR(p^a,m)$ be a unit of Type $(1)$, let $\alpha^{p^s}=\zeta_0$, and let $C=\langle(x-\alpha)^i\rangle$ be a $\gamma$-constacyclic code of length $p^s$ over $\GR(p^a, m)$. Then the following are true.

\begin{itemize}
\item If $\zeta_0=\zeta_0^{-1}$, then $C$ is a $\gamma$-constacyclic self-orthogonal code of length $p^s$ over $\GR(p^a, m)$ if and only if $  \left \lceil{\frac{ap^s}{2}}\right \rceil  \leq i \leq ap^s$.

\item If $\zeta_0 \not =\zeta_0^{-1}$, then $C$ is a $\gamma$-constacyclic self-orthogonal code of length $p^s$ over $\GR(p^a, m)$  if and only if $\left \lceil{\frac{a}{2}}\right \rceil  p^s \leq i\leq ap^s$.
\end{itemize}
}

\vskip 6pt
{\it Proof.} It follows from Proposition 3.5 that the dual of $C$ is
$$C^\perp=\langle(x-\alpha^{-1})^{ap^s-i} \rangle \subseteq \R_p(a,m,\gamma^{-1}).$$
If $C$ is self-orthogonal then we  have $|C|\le |C^{\perp}|$, which gives  $2i\ge ap^s$.

If $\zeta_0=\zeta_0^{-1}$, by Proposition~3.7, $C^\perp$ is also a $\gamma$-constacyclic code. Observing that $\alpha^{p^s}=\zeta_0=\zeta_0^{-1}=(\alpha^{-1})^{p^s}$ and by Corollary 3.8, it follows that $C^\perp=\langle(x-\alpha)^{ap^s-i} \rangle \subseteq \R_p(a,m,\gamma)$. Hence, $C$ is self-orthogonal if and only if $\langle(x-\alpha)^i\rangle \subseteq \langle(x-\alpha)^{ap^s-i}\rangle$ if and only if $\left \lceil{\frac{ap^s}{2}}\right \rceil  \leq i \leq ap^s$.

If $\zeta_0 \not = \zeta_0^{-1}$, by Proposition 2.3 and Lemma 3.4,  $ \zeta_0-\zeta_0^{-1}$ is invertible in $\GR(p^a,m)$ and $\gamma^{-1}=\zeta_0^{-1}+p\zeta^\prime_1+p^2z^\prime$. Now we consider the polynomial $x-\alpha$ in  $\R_p(a,m,\gamma^{-1})$. If $p=2$, by Proposition 3.1, we have
\begin{align*}
(x-\alpha)^{2^s}&=x^{2^s}+\alpha^{2^s}+2\alpha_s(x) \\
				&=\gamma^{-1}+\zeta_0+2\alpha_s(x)\\
				&=\zeta_0^{-1}+2\zeta^\prime_1+4z^\prime+\zeta_0+2\alpha_s(x) \\
				&=\zeta_0+\zeta_0^{-1}+2(\zeta_1^\prime+2z^\prime+\alpha_s(x)).
\end{align*}
If $p$ is odd, then
\begin{align*}
(x-\alpha)^{p^s} &= x^{p^s} + (-\alpha)^{p^s} + p(x-\alpha)\beta_s(x)\\
				&=\gamma^{-1}-\zeta_0+p(x-\alpha)\beta_s(x)\\
				&=\zeta_0^{-1}-\zeta_0+p(\zeta_1+pz^\prime+ (x-\alpha)\beta_s(x)).
\end{align*}
This gives that $(x-\alpha)^{p^s}$ is invertible in $\R_p(a,m,\gamma^{-1})$. Hence, $x-\alpha$ is also invertible in $\R_p(a,m,\gamma^{-1})$. By the division algorithm, there exist nonnegative integers $t$ and $j$, such that $i=tp^s+j$, and by Lemma 3.2, we get
$$C=\langle (x-\alpha)^i\rangle =\langle p^t(x-\alpha)^j\rangle$$
and
$$C^\perp=\langle (x-\alpha^{-1})^{ap^s-i}\rangle =\langle p^{a-t-1}(x-\alpha^{-1})^{p^s-j}\rangle.$$
If $j=0$, then $C \subseteq C^\perp $ if and only if $t \geq \left \lceil{\frac{a}{2}}\right \rceil$ if and only if $i \geq p^s \left \lceil{\frac{a}{2}}\right \rceil$.

Now we assume that $j \not = 0$. If $t<a-t-1$, then $\vert C \vert > \vert C^\perp \vert $, and hence, in this case $C$ is not self-orthogonal.
 If $t=a-t-1$  and suppose that $C \subseteq C^\perp $  then $p^t(x-\alpha)^j \in C^\perp $, which implies that $p^t \in C^\perp$, since $x-\alpha$ is invertible in $\R_p(a,m,\gamma^{-1})$ by the discussion above. Then $j=0$ and it follows that $C$ is not self-orthogonal in this case either.

If $t \geq a-t$, then
$$p^t\in \langle p^{a-t}\rangle= \langle (x-\alpha^{-1})^{p^s(a-t)}\rangle \subseteq \langle p^{a-t-1}(x-\alpha^{-1})^{p^s-j}\rangle= C^\perp.$$
Therefore, $C$ is self-orthogonal if and only if $t \geq a-t$ if and only if $i \geq p^s\left \lceil{\frac{a}{2}}\right \rceil +1$. $\square$

\vskip 6pt
\noindent {\bf Corollary 3.11.}{ Let $\gamma=\zeta_0+p\zeta_1+p^2z$ be a unit of Type $(1)$ of $\GR(p^a,m)$. Then the following are true.
\begin{itemize}
\item If $\zeta_0=\zeta_0^{-1}$, then there exists a self-dual $\gamma$-constacyclic code of length $p^s$ over $\GR(p^a,m)$ if and only if $ap$ is even. In this case, $\langle(x-\alpha)^{p^{s}a/2}\rangle$ is the unique self-dual $\gamma$-constacyclic code of length $p^s$ over $\GR(p^a,m)$.

\item If $\zeta_0 \not =\zeta_0^{-1}$, then there exists a self-dual $\gamma$-constacyclic code of length $p^s$ over $\GR(p^a,m)$ if and only if $a$ is even. In this case, $\langle p^{a/2}\rangle$ is the unique self-dual $\gamma$-constacyclic code of length $p^s$ over $\GR(p^a,m)$.

\end{itemize}
}
\vskip 6pt
{\it Proof.}{ Let $C$ be a $\gamma$-constacyclic code of length $p^s$ over $\GR(p^a,m)$, then $C=\langle (x-\alpha)^i\rangle$ and $C^\perp=\langle (x-\alpha^{-1})^{ap^s-i}\rangle$, where $ 0 \leq i \leq ap^s$. Note that $C=C^\perp$ if and only if $\vert C\vert=\vert C^\perp\vert$ and $C \subseteq C^\perp$. If $\vert C\vert=\vert C^\perp\vert$ then  $i=ap^s-i$. The rest of the proof follows from Theorem 3.10.

If $\zeta_0 \not =\zeta_0^{-1}$ and $a$ is an odd number or $\zeta_0 =\zeta_0^{-1}$ and $ap$ is odd, by Theorem 3.10, if $C$ is self-orthogonal then $ap^s-i< i$. Hence, self-dual $\gamma$-constacyclic codes do not exist in this case. $\square$

}

\vskip 6pt
\noindent {\bf Remark 3.12.}{ The $\gamma$-constacyclic codes of Type $(1)$ of $\GR(2^a,m)$, namely  $\gamma=\zeta_0+2\zeta_1+\cdots+2^{a-1}\zeta_{a-1}$ and $\zeta_0 \not = 0 \not = \zeta_1$,   are a generalization of negacyclic codes which are investigated in \cite{dinh2005}, and $\lambda$-constacyclic codes, where $\lambda$ is unit of the form $4z-1$ which are investigated in \cite{liu2016}. In fact if we take $\zeta_0=\zeta_1=\cdots=\zeta_{a-1}=1$, then $\gamma=1+2+2^2+\cdots+2^{a-1}=-1$, and in this case a $\gamma$-constacyclic code is just a negacyclic code; if we take $\zeta_1=\zeta_2=1$, in this case $\gamma=1+2+4\zeta_2+\cdots+2^{a-1}\zeta_{a-1}=-1+4z$. Hence, $\gamma$-constacyclic codes generalize both negacyclic and $\lambda$-constacyclic codes.
}

In the following, we study the structure of $\R_p(a,m,\gamma)$, where $\gamma$ is of Type $(0)$.

\vskip 6pt
\noindent {\bf Theorem 3.13.} (Cf. \cite{dinh2017}) Let $\lambda$ be a unit of a finite chain ring $R$ of characteristic $p^a$ such that there is an element $\lambda_0 \in R$ such that $\lambda_0^{p^s}= \lambda$. In $\frac{R[x]}{\langle x^{p^s}-\lambda\rangle}$, $x-\lambda_0$ is nilpotent with nilpotency index $a^{p^s}-(a-1)p^{s-1}$.

\vskip 6pt
\noindent {\bf Proposition 3.14.}{
Let $\gamma=\zeta_0+p^2z$ be a unit of Type $(0)$, and let $\alpha \in \GR(p^a,m)$ such that $\alpha^{p^s}=\zeta_0$. Then the ring $\R_p(a,m,\gamma)$ is a local ring with the unique maximal ideal $\langle x-\alpha, p\rangle$. If $z=0$, then $x-\alpha$ has nilpotency $a^{p^s}-(a-1)p^{s-1}$.
}
\vskip 6pt
{\it Proof.}{ By the same method used in Lemma 3.2, we can prove that $x-\alpha$ is noninvertible in $\R_p(a,m,\lambda)$. Let $f(x)\in \R_p(a,m,\gamma)$, then $f(x)$ can be expressed as
$$f(x)=b_0+b_1(x-\alpha)+b_2(x-\alpha)^2+\cdots+b_{p^{s-1}}(x-\alpha)^{p^{s-1}},$$
where $b_i \in \GR(p^a,m)$. Clearly, $b_1(x-\alpha)+b_2(x-\alpha)^2+\cdots+b_{p^{s-1}}(x-\alpha)^{p^{s-1}}$ is noninvertible in $\R_p(a,m,\gamma)$. Note that, $\R_p(a,m,\gamma)$ is a local ring, then $f(x)$ is noninvertible if and only if $b_0\in p\GR(p^a,m)$. Thus the ideal $\langle p,x-\alpha\rangle$ is the set of all noninvertible elements of $\R_p(a,m,\gamma)$. Hence, $\R_p(a,m,\gamma)$ is a local ring with maximal ideal $\langle p,x-\alpha\rangle$. Now suppose $p\in \langle x-\alpha\rangle$. Then there are polynomials $f_1(x)$ and $f_2(x) \in \GR(p^a,m)[x]$ such that $p=f_1(x)(x-\alpha)+f_2(x)(x^{p^s}-\gamma)$. Putting $x=\alpha$ then
\begin{align*}
p&=f_1(\alpha)(\alpha-\alpha)+f_2(\alpha)(\alpha^{p^s}-\zeta_0-p^2z)\\
 &=f_2(\alpha)(\zeta_0-\zeta_0-p^2z)=-p^2zf_2(\alpha),
\end{align*}
 which is impossible because the nilpotency index of $p$ is equal to $a$, and the nilpotency index of $p^2zf_2(\alpha)$ is less than $a$. Obviously, $x-\alpha \not \in \langle p\rangle$. Thus, $\langle p,x-\alpha\rangle$  is not a principal ideal of $\R_p(a,m,\gamma)$, which implies that $\R_p(a,m,\gamma)$ is not a chain ring. The rest of the theorem follows from Theorem 3.13.  $\square$
}

\vskip 6pt
As  mentioned earlier, the sets of Type $(0)$ and Type $(1)$ form a partition of the set of all units of $\GR(p^a,m)$ when $a \geq 2$, then from Proposition 3.14 and Theorem 3.3 we have the following theorem.
\vskip 6pt
\noindent {\bf Theorem 3.15.}{ Let $\gamma=\zeta_0+p\zeta_1+p^2z$ be a unit in $\GR(p^a,m)$, then the ring $\R_p(a,m,\gamma)$ is chain ring if and only if $\gamma$ is of Type $(1)$, i.e., $\R_p(a,m,\gamma)$ is chain ring if and only if $\zeta_1$ and $\zeta_0$ are both nonzero.
}

\section{Hamming distance}

As   mentioned in Section 3,  $\gamma$-constacyclic codes over the Galois ring $\GR(p^a,m)$  are exactly  ideals of the ring $\R_p(a,m,\gamma)$. If $\gamma$ is of Type $(1)$, then  $\gamma$-constacyclic codes are precisely the ideals $\langle (x-\alpha)^i\rangle$ of the chain ring $\R_p(a,m,\gamma)$, where $i =0, 1, \cdots, p^sa$.

In this section, we will use the structure of $\gamma$-constacyclic codes of length $p^s$ over  $\GR(p^a,m)$ to compute their Hamming distances. By Theorem 3.3 and Lemma 3.2, for $0 \leq i \leq (a-1)p^s$,
$$\langle (x-\alpha)^i\rangle \supseteq \langle (x-\alpha)^{(a-1)p^s}\rangle=\langle p^{a-1}\rangle . $$
That means the Hamming distances of the codes $\langle (x-\alpha)^i\rangle$, for $0 \leq i \leq (a-1)p^s$ are 1. For the remaining values of $i$, i.e., $(a-1)p^s\leq i \leq ap^s-1$, the main tool is the Hamming distances of all $p^m$-ary constacyclic codes of length $p^s$ over $\F_{p^m}$, that were established in \cite{dinh2010}.

\vskip 6pt
\noindent {\bf Theorem 4.1.}{ (Cf. [\cite{dinh2010}, Theorem 4.11])
Let $C$ be a $\lambda$-constacyclic code of length $p^s$ over $\F_{p^m}$, then $C=\langle(\lambda_0x+1)^i\rangle \subseteq \frac{\F_{p^m}[x]}{\langle x^{p^s}-\lambda \rangle}$, for $i \in \lbrace 0,1,\cdots,p^s\rbrace$ and $\lambda_0^{p^s}=-\lambda^{-1}$. Its Hamming distance $d(C)$ is completely determined by
$$
d(C)=\left\{
\begin{array}{ll}
1, & $if $ i=0, \\

l+2$,$ & $if $lp^{s-1}+1\leq i\leq (l+1)p^{s-1}$, where $0\leq l\leq p-2, \\

(t+1)p^{k}$,$ &
\begin{array}{l}
$if $p^{s}-p^{s-k}+(t-1)p^{s-k-1}+1\leq i\leq p^{s}-p^{s-k}+tp^{s-k-1}$,$\\
$where $1\leq t\leq p-1$, and $1\leq k\leq s-1,%
\end{array}
\\
0, & $if $i=p^{s}.

\end{array}%
\right. $$
}


\vskip 6pt
\noindent {\bf Proposition 4.2.}{ Let $\gamma=\zeta_0+p\zeta_1+p^2z \in \GR(p^a,m)$ be a unit of Type $(1)$, and let $C=\langle(x-\alpha)^i\rangle \subseteq \R_p(a, m,\gamma)$ be a $\gamma$-constacyclic code, where $p^{s}(a-1) \leq i \leq ap^s$. Then the Hamming distance of $C$ is completely determined by

$$
d(C)=\left\{
\begin{array}{ll}
1, & $if $ i=p^s(a-1), \\

l+2$,$ & $if $p^s(a-1)+lp^{s-1}+1\leq i\leq p^s(a-1)+(l+1)p^{s-1}$, where $0\leq l\leq p-2, \\

(t+1)p^{k}$,$ &
\begin{array}{l}
$if $ap^{s}-p^{s-k}+(t-1)p^{s-k-1}+1\leq i\leq ap^{s}-p^{s-k}+tp^{s-k-1}$,$\\
$where $1\leq t\leq p-1$, and $1\leq k\leq s-1,%
\end{array}
\\
0, & $if $i=ap^{s}.

\end{array}%
\right. $$
}

\vskip 6pt
{\it Proof.}{ By Lemma 3.2, $\langle(x -\alpha)^{p^s}\rangle=\langle p\rangle$ in $\R_p(a, m, \gamma)$. By the division algorithm, there exists nonnegative integer $0 \leq j < p^s$, such that $i=(a-1)p^s+j$. Therefore,
$$ \langle(x-\alpha)^i\rangle = \langle (x -\alpha)^{(a-1)p^s+j}\rangle = \langle p^{a-1}(x -\alpha)^{j}\rangle.$$

Now, the ideals $\langle p^{a-1}(x-\alpha)^i \rangle$ of $\R_p(a,m,\gamma)$ are indeed the sets of elements from the ideals $\langle (x-\alpha)^i \rangle$ of $\frac{\F_{p^m}[x]}{\langle x^{p^s}-\zeta_0\rangle}$ multiplied by $p^{a-1}$. Hence, the proof follows from Theorem 4.1.  $\square$
}

\vskip 6pt
From Theorem 3.3 and Propositions 4.2, we now have the Hamming distances of all $\gamma$-constacyclic codes of length $p^s$ over $\GR(p^a,m)$.

\vskip 6pt
\noindent {\bf Theorem 4.3.}{ Let $\gamma \in \GR(p^a,m)$ be a unit of Type $(1)$, and $C$ be a $\gamma$-constacyclic code of length $p^s$ over $\GR(p^a,m)$, i.e., $C=\langle(x-\alpha)^i\rangle \subseteq \R_p(a, m,\gamma)$, for some integer $i \in \lbrace 0,1,\cdots,p^sa\rbrace$. Then the Hamming distance of $C$ can be completely determined as follows:

$$
d(C)=\left\{
\begin{array}{ll}
1, & $if $ 0 \leq i \leq p^s(a-1), \\

l+2$,$ & $if $p^s(a-1)+lp^{s-1}+1\leq i\leq p^s(a-1)+(l+1)p^{s-1}$ where $0\leq l\leq p-2, \\

(t+1)p^{k}$,$ &
\begin{array}{l}
$if $ap^{s}-p^{s-k}+(t-1)p^{s-k-1}+1\leq i\leq ap^{s}-p^{s-k}+tp^{s-k-1}$,$\\
$where $1\leq t\leq p-1$, and $1\leq k\leq s-1,%
\end{array}
\\
0, & $if $i=ap^{s}.

\end{array}%
\right. $$

}


\section{Homogeneous distance}

\vskip 6pt
The homogeneous weight on finite rings is a generalization of the Hamming weight on finite fields and the Lee weight on the residue ring of integers modulo $4$. It was first introduced in \cite{C95} over integer residue rings, and later over finite Frobenius rings. This weight has numerous applications for codes over finite rings, such as constructing extensions of the Gray isometry to finite chain rings \cite{HHN98,HL98,GS99}, or providing a combinatorial approach to MacWilliams equivalence theorems (cf. \cite{M61,M62,Wood99}) for codes over finite Frobenius rings \cite{GS00}. In this section, we shall compute the homogeneous distance of Type $(1)$ constacyclic codes over Galois rings.

\vskip 6pt
Let $a \geq 2$, then the homogeneous weight on $\GR(p^a,m)$ is a weight function on $\GR(p^a,m)$ given as

\begin{align*}
\w_{\hm}: & \GR(p^a,m) \longrightarrow \mathbb N, \quad r \mapsto & \begin{cases}
0, & \text{ if \ $r=0$,} \\
(p^m-1)\,p^{m(a-2)}, & \text{ if \ $r \in \GR(p^a,m)\,\big\backslash\, p^{a-1}\GR(p^a,m)$,} \\
p^{m(a-1)}, & \text{ if \ $r\in p^{a-1}\GR(p^a,m)\,\big\backslash\,\{0\}$}. \\
\end{cases}
\end{align*}
The homogeneous weight of a word $x=(x_0, x_1, \cdots, x_{n-1})$ of length $n$ over $\GR(p^a,m)$ is the rational sum of the homogeneous weights of its components, i.e., $\w_{\hm}(x) = \sum\limits_{i=0}^{n-1}\w_{\hm}(x_i).$

The {\it homogeneous distance} (or minimum homogeneous weight) $d_{\hm}$ of a linear code $C$ is the minimum homogeneous weight of nonzero codewords of $C$:
$$d_{\hm}(C) = \min\big\{\w_{\hm}(x-y)\,\big|\, x,y \in C, \ x \,\not= \,y\big\} = \min\big\{\w_{\hm}(c)\,\big|\, c \in C,\ c \,\not= \,0\big\}.$$

\vskip 6pt
\noindent {\bf Theorem 5.1.} { Let $\gamma\in\GR(p^a,m)$ be a unit of Type $(1)$, and let $C$ be a $\gamma$-constacyclic code of length $p^s$ over $\GR(p^a,m)$, i.e., $C=\langle (x-\alpha)^{i} \rangle \subseteq \R_p(a,m,\gamma)$, for some integer $i \in \{0, 1, \cdots, p^sa\}$. Then the homogeneous distance $d_{\hm}(C)$ of $C$ can be completely determined:

$$
d_{\hm}(C)=\left\{
\begin{array}{ll}
0, & $if $i=ap^{s},\\
(p^m-1)\,p^{m(a-2)}, & $if $ 0 \leq i \leq p^s(a-2), \\

p^{m(a-1)}, & $if $ \leq p^s(a-2)+1 \leq i \leq p^s(a-1), \\

(l+2)p^{m(a-1)}$,$ & $if $p^s(a-1)+lp^{s-1}+1\leq i\leq p^s(a-1)+(l+1)p^{s-1}$, where $0\leq l\leq p-2, \\

(t+1)p^{m(a-1)+k}$,$ &
\begin{array}{l}
$if $ap^{s}-p^{s-k}+(t-1)p^{s-k-1}+1\leq i\leq ap^{s}-p^{s-k}+tp^{s-k-1}$,$\\
$where $1\leq t\leq p-1$, and $1\leq k\leq s-1.%
\end{array}

\end{array}%
\right. $$

}
\vskip 6pt
{\it Proof.}{ By Lemma 3.2, $\langle (x-\alpha)^{p^s} \rangle = \langle p \rangle$, and therefore $\langle (x-\alpha)^{p^sj+t} \rangle = \langle p^j(x-\alpha)^{t} \rangle$.

If $0 \leq i \leq p^s(a-2)$, we get $\langle 1 \rangle \supseteq C \supseteq \langle p^{a-2} \rangle$. Since $d_{\hm}(\langle 1 \rangle) = d_{\hm}(\langle p^{a-2} \rangle) = (p^m-1)\,p^{m(a-2)}$, $d_{\hm}(C)=(p^m-1)\,p^{m(a-2)}$.

If $p^s(a-2)+1 \leq i \leq p^s(a-1)$, then $\langle p^{a-2}(x-\alpha) \rangle \supseteq C \supseteq \langle p^{a-1} \rangle$. Clearly, $d_{\hm}(\langle p^{a-1} \rangle) = p^{m(a-1)}$ and $d_{\hm}(\langle p^{a-2}(x-\alpha) \rangle) \geq 2(p^m-1)p^{m(a-2)} \geq p^{m(a-1)}$. Thus,
$$p^{m(a-1)} \leq d_{\hm}(\langle p^{a-2}(x-\alpha) \rangle) \leq d_{\hm}(C) \leq d_{\hm}(\langle p^{a-1} \rangle) = p^{m(a-1)}.$$
This implies that $d_{\hm}(C) = p^{m(a-1)}$.

If $p^s(a-1)+1 \leq i \leq p^sa-1$, then $C = \langle p^{a-1} (x+1)^j\rangle$, $1 \leq j \leq p^s-1$. Let $c \in C$, then $c$ can be expressed as
$$ c=\sum_{i=0}^{p^s-1}p^{a-1}c_ix^i, $$
where $c_i \in \GR(p^a,m)$. Hence, $d_{\hm}(C) = d(C)p^{m(a-1)}$ and the rest of the proof follows from Theorem 4.3. $\square$

}

\section{Conclusion}
In this paper we investigated $\gamma$-constacyclic codes over $\GR(p^a,m)$ of length $p^s$. We showed that the ambient ring $\R_p(a,m,\gamma)$ is a chain ring if and only if $\gamma$ is of Type $(1)$. Moreover, if $\gamma$ is of Type $(1)$, then the complete algebraic structure, the Hamming and homogeneous weight for such $\gamma$-constacyclic codes are provided. If $\gamma$ is of Type $(0)$, the ring $\R_p(a,m,\gamma)$ is a local ring with maximal ideal $\langle p, x-\alpha\rangle$, but not a chain ring. However, the complete algebraic structure of constacyclic codes of Type $(0)$ is still unknown in general. It is interesting to give the complete algebraic structure of such kind of constacyclic codes and investigate their distances. It is also a challenge to characterize all self-dual and self-orthogonal constacyclic codes of Type $(0)$. Moreover, it will be very interesting to generalize the results in this paper to constacylic codes over finite commutative chain rings.

\vskip 16pt
\noindent {\bf Acknowledgement.} The authors are very grateful to the two anonymous reviewers and the editor for their detailed comments and suggestions that improved the quality of this paper. We are also grateful to Professor Yun Fan for helpful discussions. This work was supported by NSFC (Grant No. 11171370).


\vskip 10pt
\begin {thebibliography}{100}

\bibitem{abualrub} T. Abualrub and R. Oehmke, ``On the generators of $\mathbb{Z}_4$ cyclic codes of length $2^e$", IEEE Trans. Inf. Theory, vol. 49, no 9, 2126-2133, 2003.

\bibitem{Babu01}  N. S. Babu and K. H. Zimmermann, ``Decoding of linear codes over Galois rings", IEEE Trans. Inf. Theory, vol. 47, no 4, 1599-1603, 2001.

\bibitem{Blackford03} T. Blackford, ``Negacyclic codes over $\mathbb{Z}_4$ of even length", IEEE Trans. Inf. Theory, vol. 49, no 6, 1417-1424, 2003.

\bibitem{Blackford2003} T. Blackford, ``Cyclic codes over $\mathbb{Z}_4$ of oddly even length", Discrete Appl. Math., vol. 128, no 1, 27-46, 2003.


\bibitem{calderbank93} A. R. Calderbank, A. R. Hammons, P. V. Kumar, N. J. A. Sloane, and P. Sol\'{e}, ``A linear construction for certain Kerdock and Preparata codes",  Bull. Amer. Math. Soc., vol. 29, no 2, 218-222, 1993.

\bibitem{Calderbank95} A. R. Calderbank and N. J. A. Sloane, ``Modular and $p$-adic cyclic codes", Des. Codes Cryptogr., vol. 6, no 1, 21-35, 1995.

\bibitem{Castagnoli91} G. Castagnoli, J. L. Massey, P. A. Schoeller, and N. von Seemann, ``On repeated-root cyclic codes",  IEEE Trans. Inf. Theory, vol. 37, no 2, 337-342, 1991.

\bibitem{C95} I. Constaninescu, ``Lineare Codes \"uber Restklassenringen ganzer Zahlen und ihre Automorphismen bez\"uglich einer verallgemeinerten Hamming-Metrik", Ph.D. dissertation, Technische Universit\"at, M\"unchen, Germany, 1995.


\bibitem{dinh2005} H. Q. Dinh, ``Negacyclic codes of length $2^s$ over Galois rings", IEEE Trans. Inf. Theory, vol. 51, no 12, 4252-4262, 2005.

\bibitem{dinh2007} H. Q. Dinh, ``Complete distances of all negacyclic codes of length $2^s$ over $\mathbb{Z}_{2^a}$", IEEE Trans. Inf. Theory, vol. 53, no 1, 147-161, 2007.

\bibitem{dinh2010} H. Q. Dinh, ``Constacyclic codes of length $p^s$ over $\mathbb{F}_{p^m}+ u\mathbb{F}_{p^m}$", J. Algebra, vol. 324, no 5, 940-950, 2010.

\bibitem{liu2016} H. Q. Dinh, H. Liu, X. Liu, and S. Sriboonchitta, ``On structure and distances of some classes of repeated-root constacyclic codes over Galois rings", Finite Fields Appl., vol. 43, 86-105, 2017.

\bibitem{dinh2017} H. Q. Dinh, H. D. Nguyen, S. Sriboonchitta, and T. M. Vo, ``Repeated-root constacyclic codes of prime power lengths over finite chain rings", Finite Fields Appl., vol. 43, 22-41, 2017.


\bibitem{DL04} H. Q. Dinh and S. R. L\'opez-Permouth, ``Cyclic and negacyclic codes over finite chain rings", IEEE Trans. Inf. Theory, vol. 50, no 8, 1728-1744, 2004.

\bibitem{dougherty} S. T. Dougherty and S. Ling, ``Cyclic codes over of even length", Des. Codes Cryptogr., vol. 39, no 2, 127-153, 2006.


\bibitem{GS99} M. Greferath and S. E. Schmidt, ``Gray isometries for finite chain rings and a nonlinear ternary $(36, 3^{12}, 15)$ code", IEEE Trans. Inf. Theory, vol. 45, no 7, 2522-2524, 1999.

\bibitem{GS00} M. Greferath and S. E. Schmidt, ``Finite-ring combinatorics and MacWilliams's equivalence theorem", J. Combin. Theory, Series A, vol. 92, no 1, 17-28, 2000.

\bibitem{Hammons94} A. R. Hammons, P. V. Kumar, A. R. Calderbank, N. J. A. Sloane, and P. Sol\'e, ``The $\mathbb{Z}_4$-linearity of Kerdock, Preparata, Goethals, and related codes", IEEE Trans. Inf. Theory, vol. 40, no 2, 301-319, 1994.

\bibitem{HL98} W. Heise, T. Honold, and A. A. Nechaev, ``Weighted modules and representations of codes", In : Proceedings of the ACCT, 123-129, 1998.

\bibitem{HHN98} T. Honold and I. Landjev, ``Linearly representable codes over chain rings",  Abhandlungen aus dem mathematischen Seminar der Universit  Hamburg, vol. 69, no. 1, 187-203, 1999.

\bibitem{HP03} W. C. Huffman and V. Pless, Fundamentals of error-correcting codes, Cambridge university press, 2010.

\bibitem{M61} F. J. MacWilliams, ``Error Correcting Codes for Multiple Level Transmission", Bell System Tech. J., vol. 40, no 1, 281-308, 1961.

\bibitem{M62} F. J. MacWilliams, Combinatorial problems of elementary abelian groups, PhD. Dissertaion, Harvard University, Cambridge, MA, 1962.

\bibitem{M74} B. R. McDonald, Finite rings with identity, Marcel Dekker Incorporated, 1974.

\bibitem{Nechaev91} A. A. Nechaev, ``Kerdock code in a cyclic form",  Discrete Math. Appl., vol. 1, no 4, 365-384, 1991.

\bibitem{nedeloaia2003} C. S. Nedeloaia, ``Weight distributions of cyclic self-dual codes", IEEE Trans. Inf. Theory, vol. 49, no 6, 1582-1591, 2003.

\bibitem{PH98} V. Pless, R. A. Brualdi, and W. C. Huffman, Handbook of coding theory, Elsevier Science Inc., 1998.

\bibitem{minjia1} M. Shi, P. Sol\'e, and B. Wu, ``Cyclic codes and the weight enumerator of linear codes 
over $\mathbb{F}_2 + v\mathbb{F}_2 + v^2\mathbb{F}_2$", Appl. Comput. Math., vol. 12, no 2, 247-255, 2013.

\bibitem{minjia3} M. Shi and Y. Zhang, ``Quasi-twisted codes with constacyclic constituent codes", Finite Fields Appl., vol. 39, 159-178, 2016.

\bibitem{minjia2} M. Shi, S. Zhu, and S. Yang, ``A class of optimal $p$-ary codes from one-weight codes over $\mathbb{F}_p[u]/\langle u^m\rangle$", J. Franklin Inst., vol. 350, no 5, 729-737, 2013.

\bibitem{sobhani} R. Sobhani and M. Esmaeili, ``Cyclic and negacyclic codes over the Galois ring ${\rm GR}(p^2,m)$", Discrete Appl. Math., vol. 157, no 13, 2892-2903, 2009.

\bibitem{tang1997} L. Tang, C. B. Soh, and E. Gunawan, ``A note on the $q$-ary image of a $q^m$-ary repeated-root cyclic code", IEEE Trans. Inf. Theory, vol. 43, no 2, 732-737, 1997.

\bibitem{Lint91} J. H. van Lint, ``Repeated-root cyclic codes", IEEE Trans. Inf. Theory, vol. 37, no 2, 343-345, 1991.

\bibitem{Wan99} Z. Wan, ``Cyclic codes over Galois rings", Algebra Colloq., vol. 6, no 3, 291-304, 1999.

\bibitem{Wood99} J. A. Wood, ``Duality for modules over finite rings and applications to coding theory",  Amer. J. Math., vol. 121, no 3, 555-575, 1999.


\bibitem{zimmermann} K. H. Zimmermann, ``On generalizations of repeated-root cyclic codes", IEEE Trans. Inf. Theory, vol. 42, no 2, 641-649, 1996.



\end {thebibliography}
\end{document}